\documentstyle[prl,graphics,amssymb,multicol,latexsym,aps,epsfig]{revtex}

\title{
\vskip -100pt
{  
\begin{normalsize}
\mbox{} \hfill \\
\mbox{} \hfill ITFA-00-13\\
\mbox{} \hfill July 2000\\
\vskip  50pt
\end{normalsize}
}
Transversality of the logarithmic divergences \\
 in the
Classical Finite Temperature $SU(N)$
Self-Energy}
\author{A. Arrizabalaga, B.J. Nauta and Ch. G. van Weert}

\address{Institute for Theoretical Physics, University of Amsterdam \\
Valckenierstraat 65, 1018 XE Amsterdam\\
The Netherlands\\
}
          
\date{\today}

\begin{document}
\draft



\newcommand{\be}{\begin{equation}}
\newcommand{\ee}{\end{equation}}


\def\be{\begin{equation}}
\def\ee{\end{equation}}
\def\bea{\begin{eqnarray}}
\def\eea{\end{eqnarray}}
\def\bi{\begin{itemize}}
\def\ei{\end{itemize}}
\def\bc{\begin{center}}
\def\ec{\end{center}}
\newcommand{\ts}[1]{\mbox{\scriptsize {#1}}}

\font\mybb=msbm10 at 12 pt
\def\bb#1{\hbox{\mybb#1}}
\def\Z {\bb{Z}}
\def\bbbD{{\rm \\!D}}
\def\dbar{ \makebox[.01 in]{d}}
\def\bbbr{{\rm I\!R}} 
\def\bbbn{{\rm I\!N}} 
\def\CC{{\mathchoice
{\rm C\mkern-8mu\vrule height1.45ex depth-.05ex width.05em\mkern9mu\kern-.05em}
{\rm C\mkern-8mu\vrule height1.45ex depth-.05ex width.05em\mkern9mu\kern-.05em}
{\rm C\mkern-8mu\vrule height1ex depth-.07ex width.035em\mkern9mu\kern-.035em}
{\rm C\mkern-8mu\vrule height.65ex depth-.1ex width.025em\mkern8mu\kern-.025em}}}


\maketitle

\begin{abstract}
We show that the logarithmic divergences that appear in
the classical approximation of the finite temperature $SU(N)$
self-energy are transverse. We use the Ward identities in linear
gauges and the fact that the superficial degree of divergence $d$ of a
classical diagram only depends on the number of loops $\ell$ via $d=2-\ell$.
We comment on the relevance of this result to the construction of a
low-energy effective theory beyond hard thermal loops.
\end{abstract}

\begin{multicols}{2}

\section{Introduction}

As it is well known, low-momentum excitations in a high temperature plasma behave
classically. Nevertheless, to study these excitations one cannot
simply replace the quantum thermal field theory that describes them by
a classical thermal field theory, because of the appearance of ultraviolet
Rayleigh-Jeans type divergences. One should think of a classical thermal
field theory as an effective theory at large scales (larger than the
typical interparticle distance $\sim \hbar/T$). This involves the
introduction of a cut-off $\Lambda$ in the momentum of the order of the
temperature $\Lambda \sim T/\hbar$. The resulting cut-off dependences
reflect directly the divergences of the classical theory and
indicate its different high-momentum behavior with respect to the
quantum theory. One might hope that for low-momentum correlation
functions this different behavior does not play a role so that the
physics involved at the cut-off scale $\sim T/\hbar$ is
unimportant. However, this is not so as the high-momentum modes affect, through
interactions, also the low-momentum sector of the theory in an
essential way \cite{ASY}. \\
\indent A general strategy to improve the classical theory is to include the dominant
quantum contributions from the high-momentum (hard)
modes. In this context it is important to understand the classical divergences, since they correspond to the
dominant hard-mode contributions in the quantum theory. For instance,
the linear divergences in classical non-abelian gauge theories have a one-to-one
correspondence to the well-known hard thermal loops
(HTLs) \cite{anvw,BMS}. The fact that hard thermal loops have to be
included in an effective theory for the soft (low-momentum) modes has
been known since the work of Braaten and
Pisarski \cite{braatenpisarski}. These hard thermal loops have the
following remarkable properties, namely, they are gauge invariant, they satisfy abelian-like Ward
identities \cite{FT} and they allow a kinetic formulation in terms of a
Vlasov equation \cite{BI}.\\
\indent Besides linear divergences, classical theories contain also
logarithmic divergences starting at two-loop. (The explanation of how loops arise in a
classical theory can be found in \cite{aartssmit}.) Higher-loop
diagrams are superficially finite, although they may contain linear or
logarithmic divergences in the form of one- or two-loop subdiagrams,
respectively. Thus, we have two essential kinds of classical divergences,
linear and logarithmic.\\
\indent The relevance and the physical significance of the
linear divergences (the HTLs) has been extensively studied, in particular,
an effective action which incorporates them has been developed (see
\cite{tw} or \cite{ft2} for example). \\
\indent A further step towards an effective theory beyond hard thermal loops
would then be to include the logarithmic divergences (``log-divergences'') into the
effective action. At this point some natural questions arise, namely: Do the log-divergences have the same or similar properties as the linear divergences, namely the hard
thermal loops? What are the quantum contributions corresponding to
the log divergences?\\
\indent In any case, an extra term in the effective action containing the physics
of the logarithmic divergences would enter the equations of motion of the
fields as
a current, which must be conserved for the effective theory to be
consistent. In this letter we show that the logarithmically divergent part of the classical $SU(N)$ gauge
self-energy is transverse, therefore providing current conservation at
the level of two-point functions in the equations of motion. For that we will use the Ward identities for the
self-energy at finite temperature and the results of \cite{anvw}, where
it was argued that the degree of divergence $d$ of a classical diagram
with $\ell$ loops is $d=2-\ell$ and also that there are no log divergences
at one-loop.

\section{Transversality of Logarithmic divergences}

Gauge invariance provides us with identities among different Green
functions, the so-called Ward identities. In $SU(N)$ gauge theories the
Ward identitiy for the full retarded propagator
$D^{\mbox{\scriptsize{R, full}}}_{\mu\nu}(K) $ in the covariant gauge ($\partial_{\mu}A^{\mu}=0$) is

\be
K^{\mu}K^{\nu}D^{\mbox{\scriptsize{R, full}}}_{\mu\nu}=-\alpha 
\nonumber\\
\ee

\noindent and in general, in linear gauges ($F_{\mu}A^{\mu}=0$) it is given by

\be
F^{\mu}F^{\nu}D^{\mbox{\scriptsize{R, full}}}_{\mu\nu}=-\alpha ,
\label{wardidentities}
\ee

\noindent where $\alpha$ is the gauge
parameter in both cases. The identity is the same at $T=0$ and $T \neq 0$
\cite{Weldon}\\
\indent The identities can be written for the self energy $\Pi_{\mu\nu}$ using the relation
$(D_{\mbox{\scriptsize full}})^{-1}=(D_{0})^{-1}- \Pi$. Now, a difference
arises for the two cases $T=0$ and $T \neq 0$. This difference can be
better seen in the covariant gauge. At zero temperature,
$\Pi_{\mu\nu}$ must be a linear combination of the two available
tensors $g_{\mu\nu}$ and $K_{\mu}K_{\nu}$, which immediately leads to the result that
the self-energy is transverse, namely
$K^{\mu}\Pi_{\mu\nu}=0$. However, at finite temperature the difference
arises due to the presence of a heat bath with four-velocity
$U^{\mu}$. From (\ref{wardidentities}) and the presence of $U^{\mu}$ the Lorentz
structure of $D_{\mbox{\scriptsize{R, full}}}^{\mu\nu}$ allows different independent tensors combination of
both $K_{\mu}$ and $U_{\mu}$, for instance $g_{\mu\nu}$,
$K_{\mu}K_{\nu}$, $U_{\mu}U_{\nu}$ and
$K_{\mu}U_{\nu}+U_{\mu}K_{\nu}$. More convenient are the dimensionless
tensors $A_{\mu\nu}$, $B_{\mu\nu}$, $C_{\mu\nu}$ and $D_{\mu\nu}$
detailed in \cite{GPY,KK,landsman,KKR,Weldon}. Here we will use the
fact that both $A_{\mu\nu}$ and $B_{\mu\nu}$ are transverse to the
four-momentum $K^{\mu}$,
i.e. $K^{\mu}A_{\mu\nu}=0$ and $K^{\mu}B_{\mu\nu}=0$, whereas
$C_{\mu\nu}$ and $D_{\mu\nu}$ are not. Moreover, $A_{\mu\nu}$ and
$B_{\mu\nu}$ are respectively transverse and longitudinal to the
spatial momentum $\mathbf{k}$. With these tensors the self-energy can be written as

\be
\Pi_{\mu\nu}=\Pi_{T}A_{\mu\nu}+\Pi_{L}B_{\mu\nu}+\Pi_{C}C_{\mu\nu}+\Pi_{D}D_{\mu\nu},
\label{decomposition}
\ee

where $\Pi_{T}$ and $\Pi_{L}$ denote the four-momentum transverse
components (transverse and longitudinal to the spatial momentum
respectively) and $\Pi_{C}$ and $\Pi_{D}$ are the non-transverse components. 
This decomposition of the self-energy in the above basis of tensors is
not only valid for the covariant gauge, but also for the temporal and
Coulomb gauges and, in general, for linear gauges that do not break rotational invariance
\cite{Weldon}. \\
\indent From the Ward identity (\ref{wardidentities}) and the
decomposition (\ref{decomposition}) above one can derive a relation between the
different transverse and non-transverse components of the
self-energy \cite{Weldon}:

\be
[\Pi_{C}(K)]^{2}=[K^2 + \Pi_{L}(K)] \Pi_{D}(K).
\label{weldonformula}
\ee

We note that this result does not imply that the self-energy is
transverse. Indeed, it is well known that already at one-loop the
self-energy is not transverse \cite{KKM}. However, remarkably, the
hard thermal loop part of the self-energy is, i.e.
$K^{\mu}\Pi^{\mbox{\scriptsize HTL}}_{\mu\nu}=0$. This is due to the fact that
HTLs satisfy abelian-like ward identities \cite{FT}.\\
\indent The Ward identity (\ref{weldonformula}) will be a starting point in our discussion on the
transversality of the divergent parts of the classical self-energy.

\subsection{Linear divergences}

Let us now consider the classical approximation of $SU(N)$ gauge
theory, which is obtained by taking the $\hbar \rightarrow 0$ limit of
the quantum theory. The classical theory is expected to be a good
approximation at low energies because the classical and low-energy
limit of the Bose-Einstein distribution function $n(\omega_{\mathbf{k}})$ yield the same result:

\be
n(\omega_{\mathbf{k}})=\frac{1}{e^{\hbar\omega_{\mathbf{k}}/T}-1} \; \;
\longrightarrow \; \; \;
n_{\ts{cl}}(\omega_{\mathbf{k}}) \equiv \frac{T}{\omega_{\mathbf{k}} \hbar},
\label{distribution}
\ee

where $\omega_{\mathbf{k}}=|\mathbf{k}|$ is the frequency at
wavenumber $\mathbf{k}$. As mentioned in the introduction, the
classical theory as an effective theory requires the introduction of a
cut-off $\Lambda$ which appears in the calculation of diagrams,
reflecting directly the divergences of the theory.

The linearly divergent terms in the classical theory correspond to the HTLs in the quantum
theory \cite{anvw}. This is so because the HTLs are proportional to
the $\omega_{\ts{pl}}^{2}$, where
$\omega_{\ts{pl}} \sim gT\hbar^{-1/2}$ is the plasmon
frequency. Thus, the HTLs behave as $1/\hbar$, which become the linear divergences in the
classical theory as we take $\hbar \rightarrow 0$.

The fact that HTLs are transverse indicates that the linearly divergent terms should also be so. This can
be checked by making use of (\ref{weldonformula}).
Consider the case $K^2=K_{\mu}K^{\mu} \neq 0$ and let us start at one-loop. Since
$\Pi_{C}=0$ at tree level, it begins at order $O(g^2)$ (one-loop). Now, from
(\ref{weldonformula}) we notice that $\Pi_{D}$ should start at
$O(g^4)$ (two-loops). The two-loop contribution $\Pi_{D}^{[2]}$ (the superscript denotes loop order) is
superficially log divergent, containing at most a linear
subdivergence. Hence by (\ref{weldonformula}) we see that the one-loop
contribution $\Pi_{C}$ cannot have a linear divergence. Thus, at
one-loop both $\Pi_{C}$ and $\Pi_{D}$ vanish, and therefore, 

\be
K^{\mu}\Pi^{\mbox{\scriptsize [1], lin}}_{\mu\nu}=0,
\ee
as we expected from our considerations of hard thermal loops above. In
fact, this is another way of showing that HTLs are transverse.

\subsection{Logarithmic divergences} At one-loop there are no logarithmic
divergences \cite{anvw}, so we consider the case of two-loop, i.e. $O(g^4)$. We first split
the two-loop self-energy component $\Pi_{D}^{\ts{[2]}}$ in a logarithmically divergent part, a
part that may contain a linear subdivergence and a finite part:

\be 
\Pi_{D}^{\mbox{\scriptsize [2]}}=\Pi_{D}^{\mbox{\scriptsize [2],
log}}+\Pi_{D}^{\mbox{\scriptsize [2], sublin}}+\Pi_{D}^{\mbox{\scriptsize [2], fin}}.
\label{split1}
\ee

We insert this expression into the Ward identity (\ref{weldonformula}). The
terms in $\Pi_{C}$ at the right-hand side that match with the $O(g^4)$
at the left-hand side are those corresponding to one-loop, which does
not have logarithmic divergences, and therefore 

\be
\Pi_{D}^{\mbox{\scriptsize [2], log}}=0.
\label{dlog}
\ee

We saw already that $\Pi_{C}$ does not contain a linear divergence,
thus also
\be
\Pi_{D}^{\mbox{\scriptsize [2], sublin}}=0.
\ee

Next, we consider $\Pi_{C}$. Analogously to (\ref{split1}), we split it
in terms of the different types of divergences

\be 
\Pi_{C}^{\mbox{\scriptsize [2]}}=\Pi_{C}^{\mbox{\scriptsize [2],
log}}+\Pi_{C}^{\mbox{\scriptsize [2], sublin}}+\Pi_{C}^{\mbox{\scriptsize [2], fin}}.
\label{split2}
\ee

We use the Ward identity (\ref{weldonformula}) at $O(g^8)$,  which we
may write as

\bea
\left( \Pi_{C}^{\mbox{\scriptsize [2],
log}}+\Pi_{C}^{\mbox{\scriptsize [2],
sublin}}+\Pi_{C}^{\mbox{\scriptsize [2], fin}}\right) ^2
+2\Pi_{C}^{\mbox{\scriptsize [1], fin}}\Pi_{C}^{\mbox{\scriptsize
[3]}} \nonumber \\
= K^2 \Pi_{D}^{\mbox{\scriptsize
[4]}}+\Pi_{L}^{\mbox{\scriptsize [1]}}\Pi_{D}^{\mbox{\scriptsize
[3]}}+\Pi_{L}^{\mbox{\scriptsize [2]}}\Pi_{D}^{\mbox{\scriptsize
[2]}}.
\label{ten}
\eea

We now focus on the terms that could lead to a logarithmic
divergence. We keep from (\ref{ten}) terms proportional to
$(\rm{log}\,\Lambda)^{2}$. This results in

\be
\left (\Pi_{C}^{\mbox{\scriptsize [2],
log}}\right) ^2
+2\Pi_{C}^{\mbox{\scriptsize [1], fin}}\Pi_{C}^{\mbox{\scriptsize
[3]}}= K^2 \Pi_{D}^{\mbox{\scriptsize
[4]}}+\Pi_{L}^{\mbox{\scriptsize [1]}}\Pi_{D}^{\mbox{\scriptsize [3]}}.
\label{doublelog}
\ee

As a consequence of (\ref{dlog}),
$\Pi_{D}^{\mbox{\scriptsize{[2]}}}$ does not contain a
logdivergence, therefore the last term on the r.h.s. of
(\ref{ten}) does not contribute to (\ref{doublelog}). Let
us consider the products of one- and three-loop contributions. Since at
one-loop there are no log-divergences, the three-loop diagrams
must contain a double log-divergence for these products to
contribute. Now, schematically, the expression for a three-loop
diagram is

\be
\Pi^{\ts{[3]}}(P)=g^6 T^3 \int {d^{4}K \over
(2\pi)^4}{d^{4}K' \over (2\pi)^4}{d^{4}K'' \over (2\pi)^4}
f^{\ts{[3]}}(K,K',K'', P),
\ee

\noindent where $K$, $K'$ and $K''$ are the internal momenta, $P$ is the
external momentum, $g$ the coupling constant and $T$ the temperature.
The result after the integration over any two arbitrary internal
momenta can be regarded as either an expression for two disjunct one-loop
diagrams or a two-loop diagram, with external lines depending on the other momenta. Consider
for example the three-loop diagram in figure 1.

\begin{figure}
\begin{center}
\resizebox{8 cm}{3 cm}{\includegraphics{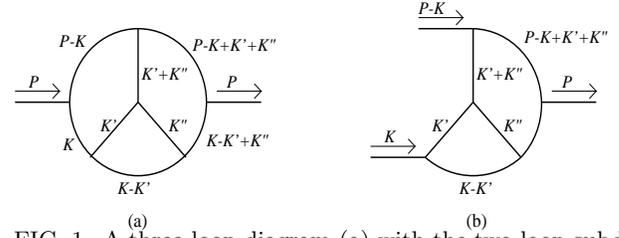}}
\caption{A three-loop diagram (a) with the two-loop subdiagram (b)}
\label{picture}
\end{center}
\end{figure}

In the case that the integration over two arbitrary internal momenta ($K'$ and $K''$ in Fig. 1.)
 corresponds to a two-loop diagram (as in Fig. 1b.), it can at most
give a single logarithmic divergence, which we denote as log$\, \Lambda$. When
it does, the integration over the remaining momentum
($K$ in Fig. 1) cannot give an extra log$\, \Lambda$, since the
superficial degree of divergence of the total diagram is
$d=2-\ell =-1$. In the case that the integration over $K'$ and $K''$
does not give a logdivergence, the integration over $K$ could
still lead to one log$\, \Lambda$. Hence, a three-loop diagram can at
most give a single log-divergence. Therefore, the product of
one- and three-loop diagrams cannot contribute to (\ref{doublelog}).\\
\indent The above argument can be repeated for the four-loop contribution to
the self-energy. In this case there are four internal momenta. The result after integration over any three given internal momenta can
be regarded as the expression for a disjunct two and one-loop diagram
or three disjunct one-loop diagrams. Therefore it can at most give a
single logarithmic divergence, and since a four-loop contribution to
the self-energy is finite, the remaining integration cannot give an
extra log-divergence and as in the case above, $\Pi^{\ts{[4]}}$
cannot contribute to (\ref{doublelog}). Thus, we find from
(\ref{doublelog}) that

\be
\Pi_{C}^{\ts{[2], log}}=0.
\ee

Note that we cannot say that the two-loop contribution to $\Pi_{C}$
containing a linear divergence from one-loop subdiagrams equals zero,
as we could for $\Pi_{D}$. \\
\indent Since both $\Pi_{C}^{\ts{[2]}}$ and $\Pi_{D}^{\ts{[2]}}$
vanish , then we conclude that the logarithmically divergent part of the
two-loop classical self-energy is transverse

\be
K^{\mu}\Pi_{\mu\nu}^{\ts{[2], log}}=0.
\label{result}
\ee

Our argument does not hold for three- or higher-loop log-divergences. However, since those diagrams are superficially finite, the divergences
can only come through two-loop subdiagrams and in general, we do not
expect the whole contribution to the self-energy to be transverse. The
important point is that, as we mentioned in the introduction, the
essential divergences that appear in the classical theory are linear
(at one-loop) and logarithmic (at two-loop) and they are both transverse.


\section{Conclusions and outlook}

Here we showed that the logarithmically divergent parts of the classical finite
temperature $SU(N)$
self-energy are transverse. This also holds for the hard thermal
loops, which correspond to classical linear divergences. Therefore we
see that all divergences appearing in the classical self-energy are
transverse. We would like to comment on the importance of this
result regarding the construction of a low-energy effective theory beyond
the HTL approximation. 

Consider for instance the effective action
which results from integrating out the hard modes with momenta $P >
\mu$, with $\mu$ an intermediate scale such that $\omega_{\rm{pl}}< \mu
 < T/ \hbar$. In a $\hbar$ or high $T$ expansion the
effective action for the soft modes would schematically be written as

\bea
\Gamma_{\ts{eff}}=g^{2}T
\left( \frac{T}{\hbar}- c_{1}\mu
\right) \overline{\Gamma}_{\ts{HTL}} 
+ \left( g^{2} T
\right) ^{2} \mbox{log}
\left( \frac{c_{2}T}{\hbar\mu} \right)
\overline{\Gamma}_{\ts{log}} \nonumber \\
+ S_{\ts{cl}} +
O\left( g^{2}\hbar, \frac{\omega_{\ts{pl}}}{\mu}, \frac{\hbar\mu}{T} \right),
\label{effectiveaction}
\eea

where $c_{1}$ and $c_{2}$ are constants that depend on the
regularization scheme. The first term in the expansion, which corresponds to the HTLs, is
proportional to $\hbar^{-1}$, being thus linearly divergent in the
classical limit $\hbar \rightarrow 0$. The second term is proportional
to log$\,$($T/\hbar$) and so corresponds to the logarithmic 
divergences in the classical theory. The third term is the classical action and the other
terms in the expansion are unimportant contributions in either a high
$T$ or classical regime. Thus we see that in a $\hbar$ or
high-temperature expansion the next-to-leading order terms are given
by the classical log-divergences. 

A consistent scheme to include hard-mode contributions beyond HTLs in
the classical theory seems to require the inclusion of terms that
diverge in the classical limit. An effective action of the form
(\ref{effectiveaction}) gives rise to currents in the equations of
motion for the classical $SU(N)$ gauge field

\be
\delta_{A_{\mu}} S_{\ts{cl}} = j_{\ts{HTL}}^{\mu}+j_{\ts{log}}^{\mu},
\ee
 
where $j_{\ts{HTL}}^{\mu} \sim \delta_{A_{\mu}}\overline{\Gamma}_{\ts{HTL}}$ and
$j_{\ts{log}}^{\mu} \sim \delta_{A_{\mu}}\overline{\Gamma}_{\ts{log}}$. The
current $j_{\ts{HTL}}^{\mu}$ generated by the HTLs is conserved. For
consistency, it is necessary that the current $j_{\ts{log}}^{\mu}$ generated by the
log-divergences is also conserved. Our result
(\ref{result}) shows that the logarithmic divergent part of the
self-energy is transverse, which implies that $j_{\ts{log}}^{\mu}$ is indeed
conserved. We stress that this is a special property at finite
temperature that should not generally be expected, and in fact, this
result encourages the study of the classical logarithmic divergences
towards the construction of a feasible low-energy effective theory beyond hard
thermal loops.

\end{multicols}

\end{document}